


\catcode`@=11


\def\pri{^\prime}


\def\titlepage{\FRONTPAGE\paperstyle\ifPhysRev\PH@SR@V\fi
   \ifp@bblock\p@bblock\fi}

\def\p@bblock{\begingroup \tabskip=\hsize minus \hsize
   \baselineskip=1.5\ht\strutbox \topspace-2\baselineskip
   \halign to\hsize{\strut ##\hfil\tabskip=0pt\crcr
   \the\Pubnum\cr \the\date\cr \the\pf\cr \the\pubmemo\cr}\endgroup}

\Pubnum={KUCP-\the\pubnum}
\pubnum={00}
\date={\monthname,\ \number\year}
\newtoks\pf
\pf={T/TE/AS}
\def\datenum{\number\month .\number\day}
\newtoks\pubmemo
\pubmemo={\it Ver. \datenum}

\def\abstract{\vskip\frontpageskip\centerline{\twelverm ABSTRACT}
              \vskip\headskip }
\def\author#1{\vskip 0.3cm \titlestyle{\twelvecp #1}\nobreak}


\date={\monthname \ \number\day,  \number\year}

\catcode`@=12

\catcode`\@=11 
%
%
\newskip\frontpageskip
\newtoks\Pubnum   \let\pubnum=\Pubnum
\newtoks\Pubnumn  
\newtoks\Pubtype  \let\pubtype=\Pubtype
\newif\ifp@bblock  \p@bblocktrue
\frontpageskip=12pt plus .5fil minus 2pt
\Pubtype={}
\Pubnum={}
\Pubnumn={KUCP-??}
\def\p@bblock{\begingroup \tabskip=\hsize minus \hsize
   \baselineskip=1.5\ht\strutbox \topspace-2\baselineskip
   \halign to\hsize{\strut ##\hfil\tabskip=0pt\crcr
       \the\Pubnum\crcr
       \the\date\crcr\the\pubtype\crcr}\endgroup}
\def\abstract{\par\dimen@=\prevdepth \hrule height\z@ \prevdepth=\dimen@
   \vskip\frontpageskip\centerline{ABSTRACT}\vskip\headskip }
\catcode`\@=12 
\def\kyoto{
{\sl
\centerline{Department of Physics, College of Liberal Arts and Sciences}
\centerline{Kyoto University, Yoshida, Kyoto 606, Japan}}}
%

%

      \def\kk{{k \kern-2.041mm k \kern-2.041mm k}}
      \def\NN{{N \kern-4.47mm N \kern-4.47mm N}}

 \pubnum{KUCP-42}
 \date{December 1991}
 \pubtype{T}
 \titlepage
 \title{Coulomb Gas Representations and Screening Operators of
    the  $\NN$=4 Superconformal Algebras}
      \author{Satoshi Matsuda\foot{Work supported in part by
         the Grant-in-Aid for Scientific Research from the Ministry
         of Education, Science and Culture (No.02640227).}}
\vskip 0.3cm
 \kyoto

\vskip 1.5cm

 \abstract

The Coulomb gas representations are presented for the
${\rm SU(2)}$$_k$-extended $N$=4 superconformal algebras,
incorporating the Feigin-Fuchs representation of the\break
${\rm SU(2)}$$_k$ Kac-Moody algebra with {\sl arbitrary} level $k$.
Then the long-standing problem of identifying
the whole set of charge-screening operators for the
$N$=4 superconformal algebras
is solved and their explicit expressions are given.
The method of achieving a rigorous proof of
the $N$=4 Kac determinant formulae
following  Kato and Matsuda is suggested.
The complete proof for them will be given elsewhere.
Our results for the screening operators also provide the basis for studying
the BRST formalism of the $N$=4 superconformal algebras
${\sl {\grave a}\ la}$ Felder.

\vskip 1.8cm
\vskip .8cm
\endpage

\def\NP{ Nucl. Phys.}

\def\PRL{ Phys. Rev. Lett.}
\def\PL{ Phys. Lett. }
\def\PTP{ Progr. Theor. Phys.}
\def\CMP{ Commun. Math. Phys.}
\def\MPL{ Mod. Phys. Lett.}
\def\IJMP{ Int. J. Mod. Phys.}

\sequentialequations

\doublespace

\PHYSREV


\REF\belavin{A.~A.~Belavin, A.~M.~Polyakov and A.~B.~Zamolodchikov
 \journal\NP &B241 (87) 333.}
\REF\ademollo{M.~Ademollo {\it et al.} \journal\NP &B114 (76) 297.}
\REF\kac{V.~G.~Kac, {\it Lecture Notes in Phys.} {\bf 94} (1979) 441.}
\REF\fuchs{B.~L.~Feigin and D.~B.~Fuchs, {\it Funct. Anal. and Appl.}
 {\bf 16} (1982) 114; {\bf 17} (1983) 241.}
\REF\meurman{A.~Meurman and A.~Rocha-Caridi \journal\CMP &107 (86) 263.}
\REF\friedan{D.~Friedan, Z.~Qiu and S.~Shenker \journal\PRL &52 (84) 1575;
   {\it Phys. Lett.} {\bf 151B} (1985) 37.}
\REF\boucher{W.~Boucher, D. Friedan and A.~Kent \journal\PL &172B (86) 316.}
\REF\kmcon{M.~Kato and S.~Matsuda \journal\PL &172B (86) 216.}
\REF\kmkac{M.~Kato and S.~Matsuda \journal\PTP &78 (87) 158.}
\REF\kmnul{M.~Kato and S.~Matsuda \journal\PL &184B (87) 184.}
\REF\kmr{M.~Kato and S.~Matsuda, {\it in} Advanced Studies in Pure
  Mathematics {\bf 16}: Conformal field theory and solvable lattice models,
  ed. M.~Jimbo, T.~Miwa and A.~Tsuchiya (Nagoya University, 1988).}
\REF\feigin{B.~Feigin and D.~Fuchs, {\it in } Representaions of
infinite-dimensional Lie groups and Lie algebras (Gordon and Breach, 1986).}
\REF\dotsenko{VI.~S.~Dotsenko and V.~A.~Fateev  \journal\NP &B240 {\rm[FS12]}
 (84) 312; {\bf B251}\ \ [FS13]\ \ (1985)\ \ 691.}
\REF\felder{G.~Felder \journal \NP &B317 (89) 215.}
\REF\kent{A.~Kent and H.~Riggs \journal \PL &198B (87) 491.}
\REF\eguchi{T.~Eguchi and A.~Taormina \journal\PL &196B (87) 75;
{\bf B200} (1988) 315.}
\REF\taormina{T.~Eguchi and A.~Taormina \journal\PL &210B (88) 125.}
\REF\yu{M.~Yu \journal\PL &196B (87) 345;
{\it Nucl. Phys.} {\bf B294} (1987) 890.}
\REF\musf{S.~Matsuda and T.~Uematsu \journal\PL &220B (89) 413.}
\REF\musd{S.~Matsuda and T.~Uematsu \journal\MPL &A5 (90) 841.}
\REF\schwim{A.~Schwimmer and N.~Seiberg \journal\PL &184B (87) 191.}
\REF\nemesha{D.~Nemeshansky \journal\PL &224B (89) 121.}
\REF\mat{S.~Matsuda, in preparation.}
\REF\mattis{A.~Kent, A.~Mattis and H.~Riggs \journal\NP &B301 (88) 426.}
\REF\miki{K.~Miki \journal\IJMP &A5 (90) 1293.}
\REF\bersha{M.~Bershadsky \journal\PL &174B (86) 285.}
\REF\kniz{V.~G.~Knizhnik, {\it Theor. Math. Phys.}\  {\bf 66}\  (1986)\  68.}


{\it 1. Introduction.}
In the last several years there has been considerable interest
in the algebraic approach to conformal field theory.\refmark{\belavin}
The structure and variety of two dimensional conformal algebras,
their representations and physical realizations thereof
have been studied to a great extent. In particular,
the study of the supersymmetric extentions \refmark{\ademollo}
of the Virasoro algebras,
being characterized by the number, $N$, of
the supersymmetry generators in the maximal superconformal subalgebra
of the given algebra, has been
one of the central issues in this field.
Physically relevant superconformal algebras (SCA's)
with interesting free field and sigma model realizations
can only have
$N$=0, $N$=1, $N$=2 and $N$=4 (and possibly $N$=3 as the slightly
exceptional but physically relevant case).
The algebras for $N>4$ contain fields of negative conformal weight
and are presumably physically uninteresting.

Kac determinant forumulae\refmark{\kac}
are the key tool in understanding the
highest weight representations of the SCA's and have been proven for the
$N$=0, $N$=1 and $N$=2.\refmark{\kac-\kmr}
The complex contour approach for proving generically
the Kac determinant formulae
was first given by Kato and Matsuda,\refmark{\kmcon-\kmr}
where the Coulomb gas representation\refmark{\feigin}
and the screening operators\refmark{\dotsenko}
for a given SCA
play the crucial role
in  obtaining nonvanishing
null states with  {\it on-shell conditions}\refmark{\kmcon}
imposed for their existence.\refmark{\kmcon,\kmnul,\kmr}
The complex contour method of Kato and Matsuda has been very
useful, and powerful as well, in various aspects: It has successfully
been applied
in proving rigorously the $N$=2 Kac determinant formulae.\refmark{\kmnul}
The method has also been employed
by Felder\refmark{\felder} to develop the BRST formalism
for minimal models
by constructing the screened vertex operators and
the BRST charge in the complex contour form.\refmark{\kmcon}

Now, some time ago, the Kac determinant formulae of
the SU(2)$_k$-extended $N$=4 SCA's
were conjectured by Kent and Riggs\refmark{\kent} by a mixture of
analytic arguments and computational experiment.
Since then their properties and implications have been studied
extensively.\refmark{\eguchi-\yu}
The unitary representations and their superspace formulation of
the $N$=4 SCA's
have been investigated as well.\refmark{\musf-\musd}
Nevertheless, a {\it rigorous} proof of
the $N$=4 Kac determinant formulae
has been left intact partly because of the awful complexity of the problem.
With the mathematically rigorous validity of the
conjectured formulae {\it unproven} so far, the previous investigations
on the $N$=4 SCA's have been left on loose grounds
inevitably.
Thus, the rigorous proof for the formulae
is the long-standing problem
having confronted us up till now.

The purpose of the present paper is to announce that it is
now possible to
achieve the mathematically rigorous proof of
the Kac determinant formulae for the SU(2)$_k$-extended $N$=4 SCA's
in terms of the generic method of
Kato and Matsuda.\refmark{\kmcon-\kmr}
In this paper we present the Coulomb gas representations
of the $N$=4 SCA's which incorporates the Feigin-Fuchs representation
of the SU(2)$_k$ Kac-Moody algebra with {\it arbitrary} level $k$.
Then we report on our remarkable construction of the whole set of
the screening operators of the $N$=4 SCA's
which is based on our Coulomb gas representation.
The results developed here form the crucial elements
in achieving the rigorous proof of the $N$=4 Kac determinant formulae.
The key points of the proof together with the on-shell conditions for
null states will be
presented.

{\it 2. The N=4 superconformal algebras.}
Now, the structure of the SCA's is essentially specified by $N$ and
for each $N$ there is a range of possible modings for the algebra
generators parametrizing distinct algebras with this basic supersymmetry
structure.\refmark{\schwim}
To get around the complexity of the moding assignments till the last
moment, we generate the $N$=4 SCA's by operators $L(z), T^i(z), G^a(z),
\bar G_a(z)$ and a c-number central charge $c=6k$ in the form of operator
product expansions (OPE's):
$$ \eqalign{
  L(z)L(w)&\sim {3k\over (z-w)^4}+{2L(w)\over(z-w)^2}+
  {\partial_wL(w)\over z-w},
                                                                    \cr
  T^i(z)T^j(w)&\sim {{1\over2}k\delta^{ij}\over(z-w)^2}+
      {{\rm i}\epsilon^{ijk}T^k(w)\over z-w},\
   \ L(z)T^i(w)\sim {T^i(w)\over (z-w)^2}+
        {\partial_wT^i(w)\over z-w},                                \cr
  T^i(z)G^a(w)&\sim -{{1\over 2}{(\sigma)^a}_bG^b(w)\over z-w}\ ,\quad
    \qquad  T^i(z)\bar G_a(w)\sim
      {{1\over 2}\bar G_b(w){(\sigma^i)^b}_a\over z-w}\ ,           \cr
  L(z)G^a(w)&\sim {{3\over2}G^a(w)\over(z-w)^2}+
     {\partial_w G^a(w)\over z-w}\ ,\
    \ L(z)\bar G_a(w)\sim{{3\over2}\bar G_a(w)\over(z-w)^2}+
         {\partial_w \bar G_a(w)\over z-w}\ ,                         \cr
  G^a(z)G^b(w)&\sim 0\ ,\quad \bar G_a(z)\bar G_b(w)\sim 0           \cr
  G^a(z)\bar G_b(w)&\sim {4k{\delta^a}_b\over(z-w)^3}-
     {4{\delta_{ij}}{(\sigma^i)^a}_bT^j(w)\over(z-w)^2}-
     {2{\delta_{ij}}{(\sigma^i)^a}_b\partial_w T^j(w)\over z-w}+
     {2{\delta^a}_bL(w)\over z-w}                                     \cr}
   \eqn\sca$$
The superscripts\ $i$=0,$\pm$\ denote SU(2) currents in the diagonal basis,
whereas the superscripts (subscripts)\ $a$=1,2\ label
SU(2) doublet (antidoublet) representations.
The group tensors $\epsilon^{ijk}, \delta_{ij}$ and
the Pauli matrices ${(\sigma^i)^a}_b$
are defined in the diagonal basis.
For later convenience
the fermionic operators are hereby re-labeled
according to the canonical notation
of SU(2) spins, thus
the index $a$  runs over $\pm$
corresponding to the increase($+$) or decrease($-$) of their
third components by a half unit when the operators are applied to
conformal states.
We henceforth have
$G^-\equiv G^1,\  G^+\equiv  G^2,\  \bar G_+\equiv \bar G_1$ and
$\bar G_-\equiv \bar G_2$.
The delta function ${\delta^a}_b$  has the standard meaning
with the SU(2) doublet or antidoublet labels, therefore  we have
${\delta^+}_+={\delta^2}_1=0,\  {\delta^+}_-={\delta^2}_2=1$, etc..

{\it 3. Coulomb gas representation.}
Our Coulomb gas theories are represented in terms of four real bosons
$\varphi_\alpha (z)$ ($\alpha$=1,2,3,4), and
four real fermions
forming\refmark{\musf}
a pair of complex fermion doublet and  antidoublet
$\gamma^a (z), \bar\gamma_a (z)$ ($a$=1,2 or $\pm$).
The SU(2)$_{\hat k}$ Kac-Moody subalgebra
with {\it arbitrary} level $\hat k$
are formed by generators $J^i$ ($i$=0,$\pm$), and their Feigin-Fuchs
representations are  given in terms of the first three bosons
by\refmark{\nemesha}
$$\eqalign{
J^0(z)     &={\rm i}{\sqrt{\hat k\over 2}}\partial\varphi_3     \cr
J^{\pm}(z) &=\ :{{\rm i}\over \sqrt 2}
            \left( {\sqrt{\hat k+2\over 2}}\partial\varphi_1\pm
    {\rm i}{\sqrt{\hat k\over 2}}\partial\varphi_2 \right)
    e^{\pm {\rm i}{\sqrt{2\over\hat k}}(\varphi_3-{\rm i}\varphi_2)}:  \cr}
        \eqn\ff$$
The corresponding contribution of the energy-momentum tensor
is given due to the Sugawara construction as
$${1\over \hat k+2}\sum_{i,j=0,\pm}:\delta_{ij}\,J^i(z)J^j(z):
    =-{1\over 2}\sum_{\alpha=1}^3:\left(\partial\varphi_\alpha\right)^2:+
        {\rm i}{\tau\over 2}\partial^2\varphi_1
     \eqn\sug$$
where $\tau\equiv\sqrt{2\over \hat k+2}$.

Now, the total energy-momentum tensor $L(z)$
is obtained by adding the contribution from the fermion doublets and
the fourth boson to Eq.{\sug}:
$$L(z)=-{1\over 2}\sum_{\alpha=1}^4:\left(\partial\varphi_\alpha\right)^2:+
        {\rm i}{\tau\over 2}\partial^2\varphi_1-
        {\rm i}\kappa\partial^2\varphi_4+
        {1\over 2}:\left( \partial\bar\gamma\cdot\gamma-
           \bar\gamma\cdot\partial\gamma \right):
              \eqn\em$$
with the parameter value of $\kappa\equiv{\rm i}\,(\hat k+1){\tau\over 2}$
where one should note the {\rm i} factor in front.

We define the total SU(2)$_k$ Kac-Moody currents, $T^i(z)$, with
level $k\equiv\hat k +1$
by adding the fermionic contribution as
$$T^i(z)=J^i(z)+{1\over 2}:\bar\gamma\sigma^i \gamma(z):
   \eqn\ka$$
The $N$=4 supercurrents $G^a(z)$ and $\bar G_a(z)$ are then given by
$$\eqalign{
 G^a(z) &={\rm i}\gamma^a \partial\varphi_4-
               2\kappa\partial\gamma^a-
       {\rm i}\tau\delta_{ij}J^i\left( \sigma^j\gamma \right)^a+
       {\rm i}\tau:\left( \bar\gamma\cdot\gamma \right)\gamma^a:    \cr
\bar G_a(z) &={\rm i}\bar\gamma_a\partial\varphi_4-
               2\kappa \partial \bar\gamma_a+
       {\rm i}\tau\delta_{ij}J^i\left( \bar\gamma\sigma^j \right)_a-
       {\rm i}\tau :\left( \bar\gamma\cdot\gamma \right) \bar\gamma_a:  \cr}
            \eqn\sup$$

We remark that our Coulomb gas representation just presented in fact
can be confirmed to satisfy the $N$=4 SCA, Eq.{\sca}, after lengthy and
tedious caluculations
for the parameter values being summarized as follows:
$$
\mu\equiv{\rm i}\tau={\rm i}\sqrt{2\over \hat k+2}\ ,\qquad
    \kappa={\rm i}\,(\hat k+1){\tau\over 2}\ ,\qquad
    k=\hat k+1
            \eqn\para$$
with {\it arbitrary} level $k$.

{\it 4. Construction of screening operators.}
Next we proceed to the construction of screening operators.
First, the vertex operators representing primary fields
of the $N$=4 SCA's are given by
$$V(t,j,j_0,z)=\ :e^{{\rm i}t\varphi_4(z)}:V_{j,\,j_0}(z)
             \eqn\vot$$
where the vertex operator $V_{j,\,j_0}(z)$ is defined as

$$V_{j,\,j_0}(z)=\ :e^{-{\rm i}j\tau \varphi_1}
   e^{{\rm i}j_0{\sqrt{2\over \hat k}}
          \left( \varphi_3-{\rm i}\varphi_2 \right)}:
\eqn\voj$$
which gives the representations labeled by SU(2)-spin
$(j,j_0)$ of
the SU(2)$_{\hat k}$ Kac-Moody algebra
and satisfy the following OPE's:
$$
J^0(z)V_{j,\,j_0}(w)\sim {j_0\over z-w}V_{j,\,j_0}(w)\ ,\qquad
J^{\pm}(z)V_{j,\,j_0}(w)\sim{-j\pm j_0\over z-w}V_{j,\,j_0\pm1}(w)
      \eqn\jope$$

Note here that the vertex operator $V_{-j-1,\,j_0}(z)$
conjugate to $V_{j,\,j_0}(z)$ has the same conformal weight $\Delta_j$
as $V_{j,\,j_0}(z)$ does, with the definition of
$$\Delta_j={\tau^2\over 2}j(j+1)={j(j+1)\over \hat k+2}
     \eqn\jcw$$
The role of the conjugate operator has to be taken into proper account
whenever the conjugate expressions obtained by its usage
provide valid results.
Thus in our Coulomb gas representation there are, in general terms,
two ways of defining
a primary state with conformal dimension $h$ and SU(2)-spin $(j,j_0)$:
$$ \vert h,j,j_0\rangle\ \sim V(t,j,j_0,z=0)\vert 0\rangle\
                \sim V(t,-j-1,j_0,z=0)\vert 0\rangle
           \eqn\pri$$
where $\vert 0\rangle$ is the ground state and the conformal dimension $h$
is given by
$$h=h(t,j)\equiv
{t^2\over 2}+\kappa t+{\tau^2\over 2}j(j+1)
  ={1\over 2}(t+\kappa)^2+{\tau^2\over 2}\left(j+{1\over 2}\right)^2
         +{\hat k\over 4}
     \eqn\cw$$

For simplicity we present the equations here for the Neveu-Schwarz case,
but our anlysis is of course general and
the results hold for other modings as well, which will be explicit later.

At this point, following ref.{\kent}, let us introduce
the ``charge'' operator $\hat C$ whose action is defined by
$$\eqalign{
\hat C\vert h,j,j_0\rangle & =0,\quad\left[\hat C, G_a(z)\right]=G_a(z),\quad
\left[\hat C,\bar G_a(z)\right]=-\bar G_a(z),                         \cr
      &\quad\left[\hat C,L(z)\right]=0,\quad\left[\hat C,T^i(z)\right]=0
\quad    \cr}
         \eqn\cha$$
The eigenvalues $\hat C$=$\pm 1$ actually correspond to  nothing but
the up and down of the third component of
SU(2)$_{global}$ spin.\refmark{\musf}

Also, for later use, let us define here the {\it chiral} vertex operators
$V^a(t,j,j_0,z)$ and $\bar V_a(t,j,j_0,z)$ $(a=\pm)$ by the following OPE
$$G^a(z)V(t,j,j_0,w)\sim {1\over z-w}V^a(t,j,j_0,w)
      \eqn\opcv$$
and by its antidoublet relation of the similar form.
Their explicit expressions are:
$$\eqalign{
V^{\mp}(t,j,j_0,z)&=(t\mp {\rm i}\tau j_0)\gamma^{\mp}(z)V(t,j,j_0,z)
         +{\rm i}\tau(j\pm j_0)\gamma^{\pm}(z)V(t,j,j_0\mp 1,z)\ ,    \cr
\bar V_{\pm}(t,j,j_0,z)&=(t\pm{\rm i}\tau j_0)\bar\gamma_{\pm}(z)V(t,j,j_0,z)
    -{\rm i}\tau(j\mp j_0)\bar\gamma_{\mp}(z)V(t,j,j_0\pm 1,z)\ .    \cr}
        \eqn\cv$$
Note that the chiral vertex operators satisfy the following OPE relations:
$$G^{\mp}(z)V^{\mp}(t,j,j_0,w)\sim 0 ,\qquad\quad
  \bar G_{\pm}(z)\bar V_{\pm}(t,j,j_0,w)\sim 0
             \eqn\chc$$

Now in the following we shall present our construction  of the
whole set of the charge screening operators
with conformal dimension one
for the SU(2)$_k$-extended
$N$=4 SCA's:

\noindent{\it (a) Nonchiral screening operator.}
Let us start with giving the {\it nonchiral}  screening  operator $N(z)$,
which basically generates those null states with the same
isospin and ``charge "quantum numbers
as the primary highest weight state. Only the mass level $n$ is raised
relative to the primary ground state.
After a pile of calculations $N(z)$ is found to be given by
$$\eqalign{
   &N(z)=\ :\left[{ \left({\rm i}\partial\varphi_4 \right)^2
+{\rm i}\mu\partial^2\varphi_4
+\kappa(2\kappa+\mu)\left(\bar\gamma\cdot\gamma\right)^2
-(\partial\bar\gamma\cdot\gamma-\bar\gamma\cdot\partial\gamma)
                              +\tau^2\delta_{ij}J^iJ^j } \right.  \cr
         &\qquad\qquad   \left.  {
      -2\delta_{ij}(\bar\gamma\sigma^i\gamma) J^j} \right]
             V(t=-2\kappa-\mu=-{\rm i}{2\over \tau},j=0,j_0=0,z):
                      \cr
                                           }   \eqn\nscr$$
Its OPE's with the $N$=4 generators  either vanish  or turn into
total derivatives.\refmark{\kmr}  To save space, we only present here
the nontrivial relations:
$$\eqalign{
L(z)N(w)& \sim \partial_w\left[{1\over z-w}N(w)\right]    \cr
G^a(z)N(w)&\sim \partial_w\left[{1\over z-w}
  \left( \mu\partial\gamma^a+\gamma^a{\rm i}\partial\varphi_4
        -2\kappa\gamma^a(\bar\gamma\gamma)
     -{\rm i}\tau\delta_{ij}\left(\sigma^i\gamma\right)^aJ^j\right)(w)
           \right]                                        \cr}
                              \eqn\opnscr$$
The similar OPE relation holds for $\bar G_a(z)N(w)$ as well.

\noindent{\it (b) SU(2)$_{k}$ Kac-Moody screening operator.}\
Another ``charge" preserving screening operator is given
by\refmark{\nemesha}

$$J_{(\pm)}(z)=\ :J^{\pm}(z)V_{j=-1,\,j_0=\mp 1}(z):\
  =\ :{\rm i}\left({1\over \tau}\partial\varphi_1
     \pm{\rm i}\sqrt{\hat k\over 2}\partial\varphi_2\right)
       e^{{\rm i}\tau\varphi_1}:
          \eqn\kscr$$
whose OPE relations  with the $N$=4 generators  either are
trivially zero or become nontrivial total derivatives.
Only some of the nontrivial OPE's  are presented here:
$$\eqalign{
&J^{\mp}(z)J_{(\pm)}(w)\sim {1\over 2}\partial_w\left[
  {\hat k+2\over z-w}V_{-1,\mp1}(w)\right]\ ,     \cr
&G^a(z)J_{(\pm)}(w)\sim {1\over 2}\partial_w\left[
  {\hat k+2\over z-w}(\sigma^{\pm}\gamma)^a(w)V_{-1,\mp1}(w)\right]   \cr}
      \eqn\opkscr$$
with the similar OPE relation for $\bar G_a(z)J_{(\pm)}(w)$ being valid.
We remark here that we have essentially a single SU(2)$_k$ Kac-Moody
charge screening operator $J(z)$:
$$J(z)\equiv J_{(+)}(z)\sim -J_{(-)}(z)
     \eqn\j$$
since the sum $J_{(+)}(z)+J_{(-)}(z)$
is just a total derivative.

\noindent{\it (c) Chiral screening operators.}
The chiral screening operators
$V^{(\mp)}(z)$ and $\bar V_{(\pm)}(z)$
carrying  the
``charge" of $\hat C$=$\pm 1$
are given in terms of the chiral vertex operators, Eq.{\cv},
as follows:
$$\eqalign{
V^{(\mp)}(z)       & \equiv
V^{\mp}(t=-{\rm i}{\tau\over 2},j={1\over 2},j_0=\pm{1\over 2},z)
    =-V^{(\pm)}(z)\equiv \mp C(z) ,
                                                         \cr
 \bar V_{(\pm)}(z) & \equiv
  \bar V_{\pm}(t=-{\rm i}{\tau\over 2},j={1\over 2},j_0=\mp{1\over 2},z)
     =\bar V_{(\mp)}(z)\equiv \bar C(z),
                                                        \cr}
   \eqn\cscr$$
where one should note that the ``charge" conjugate expressions of the
above chiral vertex operators
identically vanish:
$$  \bar V_{\pm}(t=-{\rm i}{\tau\over 2},j={1\over 2},j_0=\pm{1\over 2},z)
    \equiv 0 ,\qquad
   V^{\mp}(t=-{\rm i}{\tau\over 2},j={1\over 2},j_0=\mp{1\over 2},z)
    \equiv 0 .
                   \eqn\vcscr$$
Here again the
OPE's of $V^{(\mp)}(z)$ and $\bar V_{(\pm)}(z)$
with the $N$=4 generators either vanish or
are expressed as total derivatives.
In particular, $V^{(\mp)}(w)
$ are annihilated by $G^a(z)$ while
$\bar V_{(\pm)}(w)$ by $\bar G_a(z)$.
We only show some of the nontrivial relations taking the form of
total derivatives:
$$\eqalign{
\bar G_{\pm}(z)V^{(\mp)}(w)  \sim 2\partial_w
    \left[{1\over z-w}V(t=-{\rm i}{\tau\over 2},j={1\over 2},
    j_0=\pm{1\over 2},w)\right] ,                                   \cr
\bar G_{\mp}(z)V^{(\mp)}(w)  \sim -2\partial_w
    \left[{1\over z-w}V(t=-{\rm i}{\tau\over 2},j={1\over 2} ,
    j_0=\mp{1\over 2},w)\right] ,                                   \cr}
        \eqn\opcscr$$
whose ``charge" conjugate OPE relations
for $G^{\mp}(z)\bar V_{(\pm)}(w)$ and $G^{\pm}(z)\bar V_{(\pm)}(w)$
also hold in  similar forms.

To summarize, these screening operators are conformally invariant
vertex operators.

{\it 5. On-shell conditions for null fields and $N$=4 Kac determinant
formulae.}
Following the complex contour method
of Kato and Matsuda,\refmark{\kmcon-\kmr}
one can construct the whole
set of singular vertex operators (SVO) representing the {\it null fields}
by the use of the above screening operators:

\noindent{\it (a) Nonchiral SVO.}  Applying the screening operator
$N(z)$ on the primary vertex operator, we obtain the nonchiral null fields as
$$\Phi_p(t,j,j_0,z)=\oint_{C_p}dz_p\,N(z_p)
\int_z^{z_p}dz_{p-1}\,N(z_{p-1})
  \cdots\int_z^{z_2}dz_1\,N(z_1)\,V(t,j,j_0,z)
      \eqn\ncnull$$
where the  complex contour $C_p$
dragging the points $z_i (i=1,2,\cdots,p-1)$
is taken to encircle the origin and $z$.
Now, following Kato and Matsuda,\refmark{\kmcon,\kmr}
the existence condition of the null field $\Phi_p(t,j,j_0,z)$ gives the
on-shell condition
$$p\left(-{\rm i}{2\over \tau}\right)t+{p(p-1)\over 2}
\left(-{\rm i}{2\over \tau}\right)^2-p+n=0
     \eqn\onnc$$
with level excitation $n=pq$ ($q$=positive integers). Therefore we have
$$t=t_{(-p,q)}\equiv {\rm i}{1\over \tau}(p-1)-{\rm i}{\tau\over 2}(q-1)
    \eqn\tpq$$
The conformal dimension $h_{p,n}^{\rm R}$ of
the Ramond state with level $n=pq$ created by the null field
$\Phi_p(t_{(-p,q)},j,j_0,z)$ is given by
$$\eqalign{
h_{p,n=pq}^{\rm R}&={1\over 2}{t_{(-p,q)}}^2+\kappa t_{(-p,q)}
       +{\tau^2\over 2}j(j+1)+{1\over 16}\times 4              \cr
       & ={1\over 2}{t_{(p,q)}}^2+\kappa t_{(p,q)}
       +{\tau^2\over 2}j(j+1)+{1\over 4} +pq          \cr
       &=h\left(t_{(p,q)},j\right)+{1\over 4}+pq                    \cr}
          \eqn\cdnc$$
This result confirms the first infinite products of facotrs
$f_{(p,q)}(h,{\hat k+1\over 2},\hat t=j+{1\over 2};\rho=\eta=0)$
in the conjectured $N$=4 Kac determinant formulae,\refmark{\kent}
where
$ \hat t$ stands for the $\oint T^0(z)\,dz$ spin of
the primary  Ramond ground state
with highest weight isospin.

\noindent{\it (b) SU(2)$_k$ Kac-Moody SVO.}
Operating the screening operator $J(z)$ on the primary vertex operator,
one gets the Kac-Moody null fields:
$$\Psi_r(t,j-r,j_0,z)= \oint_{C_r}dz_r\,J(z_r)
\int_z^{z_r}dz_{r-1}\,J(z_{r-1})
  \cdots\int_z^{z_2}dz_1\,J(z_1)\,V(t,j,j_0,z)
      \eqn\kacnull$$
Again, following ref.8, the on-shell condition is obtained as
$$r\tau(-\tau j)+{r(r-1)\over 2}\tau^2+n=0
                 \eqn\onkac$$
with level excitation $n$=$rs$ ($s$=zero or positive integers).
$n$=0 is allowed because the isospin of the primary ground state
changes anyway for $r>0$.
Thus we get
$$j-r+{1\over 2}=j_{(r,s)}+{1\over 2}
    \equiv -{1\over 2}r +{1\over \tau^2}s
            \eqn\jrs$$
One could equally use the primary vertex operator $V(t,-j-1,j_0,z)$
to construct the similar null fields $\Psi_r(t,j+r,j_0,z)$.
The on-shell condition for level $n=rs$ is then given by
$$j+r+{1\over 2}=j_{(-r,-s)}+{1\over 2}
         = {1\over 2}r -{1\over \tau^2}s
          \eqn\jrss$$
Letting integer $p$ generically stand for $s$  and $-s$, and introducing
the sign function sgn$(p)$ being defined with sgn$(0)=1$, one can
introduce the generic expression
with level $n=r\vert p\vert$:
$$ \hat t\equiv j+{\rm sgn}(p)r+{1\over 2}
           =j_{(-{\rm sgn}(p)r,-p)}+{1\over 2}
   ={1\over 2}{\rm sgn}(p)r-{1\over \tau^2}p
       \eqn\gjrp$$
where by putting $j_0=j$ in Eq.{\kacnull}
one sees that
$ \hat t$ stands for the  highest weight isospin of
the primary  Ramond ground state.
Eq.{\gjrp} exactly reproduces
the $g_{(p,q=r)}(h,{\hat k+1\over 2},\hat t;\rho=\eta=0)=0$ factors in the
conjectured $N$=4 determinant formulae.\refmark{\kent}

\noindent{\it (c) Chiral SVO.} The chiral screening operators $C(z)$ and
$\bar C(z)$ defined by Eq.{\cscr} contain the vertex operaotrs
$V(t=-{\rm i}\tau j,j={1\over 2},j_0=\pm{1\over 2},z)$ which play the
key roles in obtaining the on-shell conditions for chiral null fields.
As has been studied by Kato and Matsuda in refs.{\kmnul,\kmr}, the
chiral screening operator $C(z)$ or $\bar C(z)$
only operates once for all on the primary vertex operator. Thus we obtain
the chiral null field with ``charge" $\hat C=1$ as
$$\Theta(t,j+{1\over 2},j_0,z)=\oint_{C_w}dw\,C(w)\,V(t,j,j_0,z)
           \eqn\cnull$$
where the contour $C_w$ is taken to encircle the points $w=0,z$.
The existence condition of  the null field with level excitaion $n=\ell$
($\ell$=zero or positive intergers)
now is given by
$$\left(-{\rm i}{\tau \over 2}\right)t+\left(-{\tau\over 2}\right)
  \left(-\tau j\right)+\ell+{1\over 2}=0
     \eqn\onc$$
Therefore we have
$$t+\kappa=t_{(\ell)}\equiv -{\rm i}\tau\left(j+{1\over 2}\right)
         -{\rm i}{2\over \tau}\ell
            \eqn\tonc$$
We could equally apply $C(z)$ on the primary vertex operator
$V(t,-j-1,j_0,z)$ to construct the chiral null fields
$\Theta(t,j-{1\over 2},j_0,z)$. The on-shell condition for level
$n=\ell$ is then expressed as
$$t+\kappa=-t_{(-\ell)}= {\rm i}\tau\left(j+{1\over 2}\right)
         -{\rm i}{2\over \tau}\ell
            \eqn\tonc$$
Here we let  integer $p$ stand generically for $\ell$ and $-\ell$.
Then the generic expression is obtained:
$$t+\kappa={\rm sgn}(p)t_{(p)}=
   -{\rm sgn}(p){\rm i}\tau\left(j+{1\over 2}\right)-
     {\rm i}{2\over \tau}\vert p\vert
        \eqn\gtonc$$
The conformal dimension $\tilde h_{p,n}^{R,\epsilon=+1}$
of the Ramond state
with level $n=\vert p\vert$ created by the null field
$\Theta(t,j+{\rm sgn}(p){1\over 2},j_0,z)$ is given by
$$\eqalign{
\tilde h_{p,n=\vert p\vert}^{R,\epsilon=+1}
   &=  h\left({\rm sgn}(p)t_{(p)}-\kappa,j\right)+{1\over 16}\times 4   \cr
   &={1\over 2}\left(t_{(p)}\right)^2
      +{\tau^2\over 2}\left(j+{1\over 2}\right)^2
      +{\hat k+1\over 4}      \cr
   &={\hat k+1\over 4}-2p\hat t-{2\over \tau^2}p^2+\vert p\vert   \cr
   &=h\left({\rm sgn}(p)t_{(p)}-\kappa-{\rm i}{\tau\over 2},
         \hat t-{1\over 2}\right)+{1\over 4}
             +\vert p\vert                 \cr}
               \eqn\hc$$
where $\hat t$ is defined by
$$\hat t\equiv j+{1\over 2}+{\rm sgn}(p){1\over 2}
      \eqn\chatt$$
Putting $j_0=j$ in $\Theta(t,j+{\rm sgn}(p){1\over 2},j_0,z)$,
we find that $\hat t$ stands for the highest weight isospin of the
primary Ramond ground state.

Similarly,
constructing $\bar \Theta(t,j+{\rm sgn}(p){1\over 2},j_0,z)$ by operating
the ``charge" conjugate screening operator $\bar C(z)$ on the primary
vertex operators, one obtains the conformal dimension
$\tilde h_{p,n}^{R,\epsilon=-1}$.

All these results confirm the third infinite products of factors
$h_{p,\epsilon}(h,{\hat k+1\over 2},\hat t;\rho=\eta=0)$
in the conjectured $N$=4 determinant formulae.\refmark{\kent}

{\it 6. Conclusions.}
We have presented the Coulomb gas representations of the
SU(2)$_k$-extended $N$=4 SCA's and  identified
the whole set of screenign operators. Then we have constructed
all the null fields corresponding to
the nonchiral, SU(2)$_k$ Kac-Moody, and chiral screening operators.
According to the complex contour method of Kato and Matsuda
\refmark{\kmcon-\kmr}
we have obtained the on-shell conditions for their existence.
All these results confirm the conjectured determinant formulae of the
$N$=4 SCA's written down some time ago by Kent and Riggs.\refmark{\kent}
Our approach presented above unquestionably
proves that the mathematically rigorous proof of the $N$=4 formulae is
now at hand within the framework of Kato-Matsuda method.
The details of the proof will be presented elsewhere.\refmark{\mat}

The $N$=3 SCA's have been considered\refmark{\schwim}
and the corresponding Kac determinant formulae
have been conjectured.\refmark{\mattis}
Their rigorous proof along the line presented above is a future problem.

Let us note that the application of the complex contour method
by Kato and Matsuda
has also been attempted\refmark{\miki}
in proving the Kac determinant formulae for the
non-Lie algebraic super-extentions\refmark{\bersha,\kniz} of SCA's.

Finally we remark that our results open the door to developing the
BRST formalism for the representation theories  of  the $N$=4 SCA's,
which  Felder\refmark{\felder} attempted in the
minimal conformal field theories.
This is another subject worth studying in the future.

\vskip 0.5cm
\centerline{\twelvecp Acknowledgements}

We thank  T. Eguchi and M. Kato at University of Tokyo and T. Uematsu
at Kyoto University
for useful discussions
over the last few years while working on the
present $N$=4 problem.
Conversations with those people mentioned above and
also with H. Aoyama and H. Ooguri at Kyoto University
has motivated me a lot to finish the present project
in a completed form.


\refout
\bye